\documentclass{JHEP3}
\title{The superstring Hagedorn temperature in a pp-wave background}

\author{Gianluca Grignani\\Dipartimento di Fisica and Sezione I.N.F.N., 
Universit\`a di Perugia, Via A. Pascoli I-06123, Perugia, Italia.
\email{E-mail:grignani$@$pg.infn.it}
\thanks{Work supported in part by INFN and MURST of Italy.}}

\author{Marta Orselli\\Dipartimento di Fisica and Sezione I.N.F.N., 
Universit\`a di Perugia, Via A. Pascoli I-06123, Perugia, Italia.
\email{E-mail:orselli$@$pg.infn.it}
\thanks{Work supported in part by INFN and MURST of Italy.}}

\author{Gordon W. Semenoff\\Department of Physics and Astronomy, 
University of British Columbia,
6224 Agricultural Road, Vancouver, British Columbia V6T 1Z1 Canada.
\email{E-mail:semenoff$@$physics.ubc.ca}
\thanks{ Work supported in part by NSERC of Canada and the PIMS String Theory CRG.}}

\author{Diego Trancanelli\\Department of Physics and Astronomy, 
State University of New York at Stony Brook, Stony Brook, 
NY 11794-3800, USA.
\email{E-mail:trancane$@$grad.physics.sunysb.edu}}

\abstract{The thermodynamics of type IIB
superstring theory in the maximally supersymmetric plane wave
background is studied. 
We compute the thermodynamic partition function for non-interacting strings 
exactly and the result differs slightly from previous computations. 
We clarify some of the issues related to the Hagedorn
temperature in the limits of small and large constant RR $5$-form. 
We study the thermodynamic behavior of strings in the
case of $AdS_3 \times S^3 \times T^4$ geometries in the presence of
NS-NS and RR $3$-form backgrounds.  We also comment on the relationship
of string thermodynamics and the thermodynamic behavior of the sector of 
Yang-Mills theory which is the holographic dual of the string theory.}

\preprint{}


\begin{document} 
\def\be{\begin{equation}} 
\def\ee{\end{equation}} 
\def\bea{\begin{eqnarray}} 
\def\eea{\end{eqnarray}} 
\def\nn{\nonumber}
\def\const{{\rm const}} 
\def\v{\varphi} 
\def\s{\sigma}
\def\vcl{\varphi_{\rm cl}} 
\newcommand{\no}[1]{:\!#1\!:}
\def\la{\left\langle} 
\def\ra{\right\rangle} 
\def\d{\partial}
\def\se{S_{\rm eff}} 
\def\tr{\rm Tr}
\def\sfa{{\sf a}}
\def\sfb{{\sf b}}
\def\sfc{{\sf c}}
\def\dsfa{{\dot{\sf a}}}
\def\dsfb{{\dot{\sf b}}}
\def\dsfc{{\dot{\sf c}}}
\def\alppr{{\alpha^\prime}}
\def\bepr{{\beta^\prime}}
\def\cI{{\cal I}}
\def\cJ{{\cal J}}
\def\prp{\pi^\oplus}
\def\prm{\pi^\ominus}
\def\PP{{\cal P}}
\def\x'{\mathaccent 19 x}
\def\th'{\mathaccent 19 \theta}
\def\barth'{\mathaccent 19 {\bar{\theta}}}
\def\phx{\mathaccent 19 {\phantom{x}} }
\def\i{i}
\def\j{j}
\def\k{k}
\def\l{l}
\def\ipr{{i^\prime}}
\def\jpr{{j^\prime}}
\def\kpr{{k^\prime}}
\def\lpr{{l^\prime}}
\def\xpr{{x^\prime}}
\def\ypr{{y^\prime}}
\def\rpr{{r^\prime}}
\def\del{\partial}
\def \lr {\bibitem{ }}
\def\bd {{\bar \del}}
\def \ra {\rightarrow}
\def \vp {\varphi}
\def \tx { {\tilde} x}
\def \L {\Lambda}
\def \a {\alpha}
\def \ga {\alpha}
\def \b {\beta}
\def \k {\kappa}
\def \gb {\beta}
\def \ga {\alpha}
\def \o {\omega}
\def \gl {\lambda}
\def \gp {\phi}
\def \p {\phi}
\def \ep {\epsilon}
\def \s {\sigma}
\def\gg {\gamma}
\def \gr {\rho}
\def \r {\rho}
\def \d {\delta}\def \G {\Gamma}
\def \l {\lambda}
\def \m {\mu}
\def \g {\gamma}
\def \n {\nu}
\def \gd {\delta}
\def \gs {\sigma}
\def \gt {\theta}
\def \bgb {\bar \beta }
\def \gij {g_{ij}}
\def \Gmn {G_{\mu \nu}}
\def \fourth {{1\over 4}}
\def \third {{1\over 3}}
\def \const {{\rm const }}
\def \tf {{\tilde  f }}
\def \bbf {{\bar f}}
\def \vp {\varphi}
\def \hg {{\hat g}}
\def \B {{\hat B}}
\def \H {{\hat H}}
\def \tg {{\tilde g}}
\def \ha { { 1\over 2 }}
\def \dy   {{\dot y}}
\def \hij {{H_{ij}}}
\def \dB {{\dot B}}
\def \ov {\over}
\def \sm  { sigma model\ }
\def \sms {$\s$-models\ }
\def\Z'{\mathaccent 19 Z}
\def\mun{{\underline{m}}}
\def\nun{{\underline{n}}}
\def\kun{{\underline{k}}}
\def\pun{{\underline{p}}}
\def \adss {$AdS_5 \times S^5$\ }
\def \N {{\cal N}}
\def \lc {light-cone\ }
\def \ta { \tau}
\def \s { \sigma }
\def \sg {\sqrt {g }}
\def \te {\theta}
\def \vp {\varphi}
\def \gij {g_{ab}}
\def \xp {x^+}
\def \xm {x^-}
\def \p {\phi}
\def \vt {\theta}
\def \bx {\bar x} \def \a { \alpha}
\def \r {\rho}
\def \fourth {{1 \ov 4}}
\def \half {{1 \ov 2}}
\def \inv {^{-1}}
\def \ri {{\rm i}}
\def \D {{\cal D}}
\def \DD {{\rm D}}
\def \vr {\varrho}
\def \diag {{\rm diag}} \def \td { \tilde }
\def \tta {\td \eta}
\def \cA {{\cal A}}
\def \cB   {{\cal B}}
\def \na {\nabla}
\def \adss {$AdS_5 \times S^5$\ }
\def \N {{\cal N}}
\def \lc {light-cone\ }
\def \four{{\textstyle {1\ov 4}}}
\def \third { \textstyle {1\ov 3}}
\def\det{\hbox{det}}
\def \ci {\cite}
\def \g {\gamma}
\def \G {\Gamma}
\def \k {\kappa}
\def \l {\lambda}
\def \pw {plane-wave\ }
\def \foot {\footnote}
\def \u {x^{+}}
\def \vv {x^{-}}
\def \S {{\tilde S}}
\def \om{\omega}
\def \hv{ {\hat v}} \def \hu{ {\hat u}} \def \hi {{\hat i} }
\def \A {\bar A}
\def \F {{\cal F}}
\def \bi{\bibitem}
\def \la {\label}
\def \lc { light-cone }
\def \pw { plane-wave  }
\def \vac {|0\rangle}
\def \T {{\rm T}}
\def\cI{{\cal I}}\def\cK{{\cal K}}
\def\cJ{{\cal J}}\def\cD{{\cal D}}
\def\PP{{\cal P}}
\def\R{{\scriptscriptstyle R}}
\def\L{{\scriptscriptstyle L}}
\def \mm {\mu}
\def \vt {\theta}
\def \f {{\rm f}} \def \E {{\cal E}}
\def \H {{\rm h}}
\def \rhoh{\psi}
\def \Q {{\cal C}}
\def \D {{\cal D}}
\def \DD {{\rm D}}

\section{Introduction}

Understanding the finite temperature states of string theory is essential for many of its 
potential applications, particularly the study of black holes and early universe cosmology.
One of the fascinating features exhibited by string theories is
the exponential growth of their densities of states with energy \cite{Hagedorn:st}.
For their thermodynamics this leads to 
either a limiting, Hagedorn temperature beyond which an ensemble of strings cannot
be heated or perhaps a phase transition to a state which is better described by degrees of
freedom other than strings \cite{Huang:1970iq}.

The existence of a Hagedorn temperature is 
well-established for all consistent non-interacting string theories on Minkowski
space.  Recently, it has been noted that the noninteracting type IIB superstring can be 
solved explicitly 
on a maximally supersymmetric plane-wave background \cite{Metsaev:2001bj,Metsaev:2002re}.  
This gives a background
other than Minkowski space where some of the ideas of string theory can be tested.   
Indeed, there are now several discussions of the Hagedorn temperature on the plane-wave
background where it has been shown to be a non-trivial function of
the dimensional parameter of the background and the string scale 
\cite{PandoZayas:2002hh}-\cite{Sugawara:2003qc}.

Another fascinating aspect of string theory is the idea of holography: that string theory,
which is a theory of quantum gravity, has a dual description as a quantum field theory
living on the boundary of the background space.  In fact, the converse, 
that a gauge field theory could have a dual description as 
a string theory, is an old and important idea in particle physics \cite{Polyakov:ez}.  
There is now one explicit example of such a duality.  Maximally supersymmetric 
four-dimensional Yang-Mills theory is thought to be an exact dual of the IIB
superstring on $AdS_5\times S^5$ background~\cite{Maldacena:1997re,Gubser:1998bc,Witten:1998qj}.  
Moreover, the plane-wave background can be obtained as a Penrose limit of 
$AdS_5\times S^5$ and the analogous limit can be taken for Yang-Mills 
theory to find the Yang-Mills dual of string theory on the plane-wave background 
\cite{Berenstein:2002jq}.  
Since superstrings on the plane wave background are more tractable than on 
$AdS_5\times S^5$, many interesting aspects of this
duality can be studied explicitly.  
In particular, it gives a promising approach to understanding superstring 
interactions \cite{Spradlin:2002ar}-\cite{Roiban:2002xr}.

In this paper we will give a brief derivation of the Hagedorn
temperature on the plane wave background.  We will comment on its
relationship with previous work
~\cite{PandoZayas:2002hh,Greene:2002cd,Brower:2002zx,Sugawara:2003qc}.
In particular, we show that the Hagedorn temperature is a
monotonically increasing function of the parameter $|f|\sqrt{\alpha'}$
where $f$ is the Ramond-Ramond flux.  In the following, we also
provide some comments on the interpretation of the Hagedorn behavior
in the limit of Yang-Mills theory which is dual to the string theory.

\subsection{Hagedorn and AdS/CFT}

The AdS/CFT duality asserts that maximally supersymmetric Yang-Mills
theory in four dimensions with SU(N) gauge group is exactly dual to
type IIB superstring theory on the background space $AdS_5\times S^5$
with $N$ units of Ramond-Ramond flux~\cite{Maldacena:1997re}. The
radii of curvature of  the $AdS_5$ and $S^5$ are equal and are given by
\begin{equation}
R=\left(4\pi g_s N\right)^{1/4}\sqrt{\alpha'}
\label{couplings}
\end{equation}
The Yang-Mills and string coupling constants are related by
\begin{equation}
g_{YM}^2=4\pi g_s ~~~.
\end{equation} 

This duality gives useful information in limits where either the
Yang-Mills or string theory can be analyzed quantitatively.   Because
of difficulties in quantizing strings in background Ramond-Ramond
fields, quantitative results for the string theory on $AdS_5\times
S^5$ are only known in some limits.  The first one to be explored is
the limit where IIB string theory  coincides with classical type IIB
supergravity. This  limit is obtained by  first taking the classical
limit,  $g_s\to 0$, holding $R$ constant.  This projects onto tree
level string theory.  Then, it is necessary to take the limit of large
string tension.   This is done by putting the effective string tension
$R^2/\alpha'=\sqrt{4\pi g_sN}\to \infty$.  This isolates the lowest
energy modes of the string, which are the supergravity fields on the
$AdS_5\times S^5$ background.

On the Yang-Mills side, the first of these limits 
corresponds to taking $g_{YM}^2\to 0$ and therefore $N\to\infty$, while
holding the 't Hooft coupling  
\begin{equation}
\lambda\equiv g_{YM}^2N
\label{thooft}
\end{equation}
fixed.  This is the 't Hooft large $N$ (or planar) limit of the gauge theory.
For any process, perturbative contributions are the sum of all planar Feynman
diagrams -- those which can be drawn on a plane without crossing lines.  In planar
Yang-Mills theory, the effective coupling constant is the 't Hooft coupling
$\lambda$.  Then, the second limit, $R^2/\alpha'\to\infty$,  
corresponds to taking $\lambda\to\infty$.  This gives
strongly coupled limit of planar gauge theory.  

It is this fact, that a solvable limit of string theory 
is mapped onto a non-trivial limit of gauge theory which makes
the AdS/CFT duality so interesting.  At the same time, it makes it
difficult to check because reliable computational techniques do not have an overlapping
domain of validity.  This has limited checks of the
conjecture to objects such as two and three point functions of chiral
primary operators  \cite{Lee:1998bx}
which do not depend on the coupling constant and thus trivially
extrapolate
between weak and strong coupling, some anomalies \cite{Freedman:1998tz,Chalmers:1998xr}
where dependence on the 
coupling constant is trivial and also to the computations of expectation
values of certain Wilson loops \cite{Erickson:2000af,Drukker:2000rr}
and correlators of Wilson loops with chiral
primary operators \cite{semenoff:2001xp}.  
In the case of a circular loop the Yang-Mills
perturbation theory can be summed to all orders in planar diagrams and
extrapolated to strong coupling, finding agreement with string
theory computations in the supergravity limit.

Recently, another limit of the string and Yang-Mills 
theory has been studied \cite{Berenstein:2002jq}.  
A certain limit of the $AdS_5\times S^5$ string theory background results in
the plane-wave metric
\be 
ds^{2}= 
2dx^{+}dx^{-}-f^{2}x_{I}^{2}dx^{+}dx^{+}+dx^{I}dx^{I},\qquad
I=1,\ldots, 8,
\label{metric}
\ee
and Ramond-Ramond field with non-zero components
\be 
F_{+1234}=F_{+5678}=2f.
\label{5form}
\ee
In the above two equations and throughout this paper 
we shall use the notation of ref.\cite{Metsaev:2002re}.
In particular, $x^\pm=\frac{1}{\sqrt{2}}\left(x^9\pm x^0\right)$.

Like the supergravity limit, (\ref{metric}) is obtained from $AdS_5\times S^5$ when the curvature
is weak and the effective string tension is large, that is, $R^2/\alpha'\to\infty$.  
However, this limit is taken
asymmetrically, in a reference frame which has large angular momentum 
$J\sim R^2/\alpha'$ on
$S^5$. In this way, the limit retains a particular subset of the 
the higher level string excitations.  Those excitations are described
by quantizing the string on the background (\ref{metric},\ref{5form}).  
This background has the advantage that non-interacting string theory
can now be quantized explicitly  
\cite{Metsaev:2001bj,Metsaev:2002re}. For example, the energy spectrum of the
non-interacting string is easy to obtain.  Further, the interaction terms in
the string field Hamiltonian have been studied 
\cite{Spradlin:2002ar,Spradlin:2002rv,Pankiewicz:2002tg,Roiban:2002xr}.

In the supergravity limit, it is thought that all operators in the 
Yang-Mills theory that are not protected by supersymmetry get infinitely
large conformal
dimensions and decouple from the spectrum.  The protected operators
are just those required to match the classical field
degrees of freedom of IIB supergravity linearized about the 
$AdS_5\times S^5$ background. 
 
In the plane wave limit, it is still necessary to take $N\to\infty$ and 
$\lambda\to\infty$
and the conformal dimensions of unprotected Yang-Mills operators become
infinite.  However, now the 
operators of interest are those which the AdS/CFT duality maps onto the
string states that are
found in the limit. Those operators have infinite conformal dimension 
$\Delta\sim \sqrt{N}$ and  also infinite $U(1)\subset SO(6)$ R-charge 
$J\sim \sqrt{N}$.  Both diverge as $N\to\infty$ in such a way 
that the momenta of the corresponding string state, which are identified by \footnote{In string
theory these are defined by $p^\pm=p_\mp =\frac{1}{2\pi\alpha'}\int d\sigma \partial_\tau X^\pm$ and
have a simple form only in the light-cone gauge. In the gauge theory they are defined by the
re-scaling of $p_\pm = \frac{1}{\sqrt{2}}(\Delta\mp J)$ needed to get the plane-wave limit. }
\begin{equation}
p^-\equiv \frac{f}{\sqrt{2}}(\Delta-J)~~,~~p^+\equiv \frac{(\Delta+J)}{\sqrt{2}f R^2} ~~~,
\label{momenta}
\end{equation} 
remain finite and non-zero~\cite{Berenstein:2002jq}.

The plane-wave limit of the Yang-Mills theory
has two parameters, $g_{YM}$ and the ratio $J^2/N$ which must be held
fixed as $N\to\infty$.  Combinations of them which appear naturally in
the Yang-Mills perturbation theory are 
\begin{equation}
\lambda'= \frac{g_{YM}^2 N}{J^2}  
\label{lambdaprime}
\end{equation}
and 
\begin{equation}
g_2= \frac{J^2}{N} 
\label{g2}
\end{equation}
It was shown in refs. \cite{Kristjansen:2002bb,Constable:2002hw} that
$\lambda'$ governs the loop expansion in Yang-Mills theory and it also
fixes the distance scale in string theory.  Also, the  constant $g_2$
is the effective string coupling in that it governs the loop expansion
in string theory.  It plays the same role in Yang-Mills theory where
it weights Feynman graphs by the genus of the two dimensional surface
on which they can be drawn without crossing lines.  In light-cone
quantization it is natural to consider states which have a fixed
light-cone momentum $p^+$.  In this case, we can easily see hat $g_2$
is related to string loops   by  using the second equation in
(\ref{momenta}) to trade the Yang-Mills parameters for the pair $g_s$
and  $p^+$,  the light-cone momentum of the string,
\begin{equation}
\lambda'= \frac{2}{ \left( f\alpha'
p^+\right)^2}
\label{lambdaprime1}
\end{equation}
\begin{equation}
g_2= g_{YM}^2\frac{ \left( f\alpha' p^+\right)^2}{2}=
2\pi g_s \left( f\alpha' p^+\right)^2
\label{g21}
\end{equation}
The free string theory is obtained by putting 
$4\pi g_s=g_{YM}^2\to 0$ in conjunction with the large $N$ limit, with the combination
$\left(f\alpha' p^+\right)$ non-zero and fixed.  
This is just the limit where $\lambda'$ is held constant and $g_2$ is set to zero. 
In this limit, all quantities depend on the parameters 
$g_{YM}$ and $N$ only through the the combination $g_{YM}^2 N=\lambda$, the 
't Hooft coupling.  This means that free strings are described by 
the planar limit of Yang-Mills theory.     
It has now been checked explicitly that the spectrum of free strings is found
in the conformal dimensions $\Delta$ of certain Yang-Mills operators computed from planar 
Feynman diagrams 
\cite{Berenstein:2002jq,Gross:2002su,Santambrogio:2002sb,Minahan:2002ve}. This shows
beautiful agreement of the matching between free string states on the plane-wave backrgound 
and a certain set of operators in the planar limit of Yang-Mills theory.  
 
In this Paper, we shall examine the thermodynamic states of string theory
in this limit.  We will use the canonical ensemble.  The partition function is
the trace of the Botlzman distribution,
\begin{equation}
Z(\beta,f) \equiv e^{-\beta F(\beta,f)}={\rm Tr}\left( e^{-\beta p^0} \right) 
\label{partf}
\end{equation}
$F(\beta,f)$ is the Helmholtz free energy.
Here the trace is over all physical multi-string states.
The rest frame energy is given 
by $p^0=\frac{1}{\sqrt{2}}(p^+ -p^-)$.  

Note that, we could, as was done in ref. \cite{Greene:2002cd}, introduce a separate parameter
for $p^+$ and $p^-$ and study the theory with two parameters,
\begin{equation}
\tilde Z(a,b,f)={\rm Tr}\left( e^{-ap^+ + b p^-}\right)
\end{equation}
However, there is a symmetry of the theory which puts
$p^+\to p^+/\Lambda$, $p^-\to p^-\Lambda$ and $f\to f\Lambda $
which implies that $\tilde Z(a,b,f)$ is equal to $\tilde Z(\sqrt{ab},\sqrt{ab}, f\sqrt{a/b})$.  
Thus, by
computing $Z(\beta,f)=\tilde Z(\beta/\sqrt{2},\beta/\sqrt{2}, f)$ 
we can deduce $\tilde Z(a,b,f)$ by identifying $\beta=\sqrt{2ab}$ and 
replacing $f\to f\sqrt{a/b}$. 
 
For the free string theory, we can compute the
Helmholtz free energy in (\ref{partf}) exactly.  This should then coincide with
the free energy of Yang-Mills theory obtained from (\ref{partf}) by taking 
a trace over Yang-Mills states with the momenta identified in (\ref{momenta}) and 
where the t'Hooft large $N$ limit is taken.  

In perturbation theory, the free energy would therefore be
found as the sum of all orders in planar connected vacuum
Feynman diagrams.  However, at each order, these diagrams are proportional to $N^2$ and 
therefore diverge in the large $N$ limit.  On the other hand, 
the string theory free energy which we compute is not of order $N^2$, instead it is of order one.  
The reason for this discrepancy is 
that perturbation theory describes the de-confined phase of the gauge theory where the number of
physical degrees of freedom is indeed of order $N^2$, and is only valid
if the temperature is greater than the de-confinement transition temperature.  That is not
the regime described by free strings which rather exist only in the confined phase, found at
temperatures below the deconfinement transition and where the number of degrees of freedom is not
of order $N^2$ at large $N$, but is of order the number of color singlet operators which, at a given
energy, is
roughly constant with $N$.  In fact, it is reasonable to identify
the Hagedorn temperature, at which a description of the theory by free strings ceases to be
meaningful, as the de-confinement transition temperature \cite{Atick:1988si}.

At this point, as clarification,  
we should note that this conformally invariant Yang-Mills theory when it is quantized
on $R^3\times R^1$ does not have a confining phase.  It is always in a conformally invariant deconfined 
phase with a Coulomb-like force law for gauge theory interactions.  
However, the correct dual of the superstring is Yang-Mills theory with radial quantization, that is,
it should be quantized on the  space $S^3\times R^1$ which can be obtianed from $R^3\times R^1$ by
a conformal transformation.  It is the energy on the space $S^3\times R^1$ 
which is dual to the string energy and is in fact given by the conformal dimension 
$\Delta$ of operators of Yang-Mills theory on the original space 
$R^3\times R^1$.  From this point of view, 
the discreteness of the spectrum of $\Delta$ comes from the fact that $S^3$ has finite volume.  
Further, when $\Delta$ is used as the Hamiltonian, 
the finite temperature Yang-Mills theory lives on the space
$S^3\times S^1$ where the time direction is Euclidean and has been periodically 
identified, $X^0\sim X^0+\beta$, with the appropriate
antiperiodic boundary condition for fermions.  

Even on this space, since the volume is finite, one does not expect a confinement-deconfinement
phase transition when $N$ is finite.  This transition could only occur at infinite $N$.
However, it is just the infinite $N$ limit that must be taken to obtain strings on the plane-wave
background.  In this limit, the Yang-Mills theory could have a phase transition corresponding
to confinement-deconfinement as the temperature  is varied.  
An order parameter for such a phase transition is the Polyakov
loop \cite{Polyakov:vu},\cite{Susskind:up}
$$
\left< {\rm Tr}{\cal P}e^{i\oint_{S^1}A} \right>
$$
which, in this adjoint gauge theory, transforms under a certain discrete large gauge symmetry
related to confinement.
There are many examples of gauge theories where this order parameter can be explicitly seen to characterize confinement 
\cite{Svetitsky:1982gs}-\cite{Grignani:1995hx}.  For example, 
the one-dimensional 
non-Abelian coulomb gas studied in refs. \cite{Semenoff:1996xg},\cite{Semenoff:1996ew}
has a deconfiment transition
only at infinite $N$, corresponding to a re-arrangement of the distribution of 
eigenvalues of the unitary 
matrix ${\cal P}e^{i\oint_{S^1}A}$, analogous to that which is
well known to occur at large $N$ in unitary matrix models \cite{Gross:he}.
If, as is suggested in ref. \cite{Atick:1988si}, the de-confinement and Hagedorn behaviors can 
be identified, the existence of a Hagedorn temperature in the string theory dual is a confirmation
of the existence of a de-confinement transition in the Yang-Mills theory, at least in the planar limit which is dual to free strings.  

We shall indeed find that, in the limit where the string coupling $g_s$ is put to zero, there is a Hagedorn
temperature for all finite values of the parameter $f$ of the background, implying
that the planar Yang-Mills theory indeed has a confinement-deconfinement
phase transition.  The string theory analysis gives the value of the transition
temperature for the gauge theory.

It is interesting to contrast the situation of the plane-wave
background to that in AdS/CFT before the plane wave limit is taken.
In the latter case, the Hagedorn spectrum for the operator $\Delta$
appears in Yang-Mills theory as the exponentially increasing
multiplicity of an infinite tower of operators which are gauge
invariant traces of local products of the fields. When $N$ is
infinite, products of all sizes are independent operators. When the
't Hooft coupling $\lambda$ is small and $\Delta$ of these operators
deviates little from the tree level values, the number of operators
with a given value of $\Delta$ can be counted~\cite{Polyakov:2001af}
and it indeed grows exponentially with increasing $\Delta$, producing
a Hagedorn spectrum for Yang-Mills theory quantized on $S^3\times
R^1$. Thus, we would expect large $N$ Yang-Mills theory to have a
Hagedorn temperature if $\lambda$ is small enough. When $\lambda$
gets large, the anomalous dimensions of operators get large and they
begin to decouple from the low-lying spectrum.

At very large $\lambda$ the dynamics is that of classical
supergravity, perhaps with stringy corrections which are suppressed by
factors of $1/\sqrt{\lambda}$.  It is known that supergravity with an
asymptotically AdS geometry  has a Hawking-Page phase transition
\cite{Hawking:1982dh} between an AdS black hole state, which can be
interpreted as the de-confined phase, and one which is AdS space with
periodic euclidean time, which can be interpreted as the confined
phase.  Indeed, the fact that the free energy of the black hole is of
order $N^2$, whereas in the  periodic AdS space it is of order one is
in line with this interpretation
\cite{Witten:1998qj,Witten:1998zw}. One could then speculate that the
Hagedorn behavior which is seen in weakly coupled planar Yang-Mills
theory evolves to the Hawking-Page transition of supergravity with
periodic Euclidean time as $\lambda$ goes from zero to infinity,  and
further that this corresponds to the de-confinement phase
transition. It is also clear that the temperature where the
Hawking-Page transition occurs is proportional to the radius of
curvature of the AdS space, $T_H\sim R/\alpha'\sim\lambda^{1/4}$ and
it actually becomes large as $\lambda\to\infty$, as we expect.

In contrast, the partition function of the limit of Yang-Mills theory which 
corresponds to the plane wave background would be the trace over states of
the exponential of the operator
\begin{equation}Z={\rm Tr} \exp \left(
-\frac {\beta^2} {\alpha'} \frac {(\Delta+J)} {2\beta f \sqrt{\lambda} }
-\beta f \frac{\Delta - J}{ 2} \right)
\label{exp}
\end{equation}
We see that the parameters indeed appear naturally in the combinations $\beta^2/\alpha'$ and $\beta f$.  

\subsubsection{large $f$}

In the limit where $f$ is large for fixed $\beta$ and $\lambda$, the
states which dominate the partition sum are those with $\Delta=J$.
They are just the single and multi-trace chiral primary operators,
${\rm Tr}(Z^J_1){\rm Tr}(Z^J_2)...{\rm Tr}(Z^J_k)$ whose conformal
dimensions $\Delta=\sum J_i=J$ are protected by supersymmetry. They
correspond to single and multi-string states where the string is in
its lowest state,  the string state which is described by  the
supersymmetric vacuum of the worldsheet sigma model.

To find the partition functions, we can think of the number of times
$J_1$ appears in the product of traces as the occupation number
$n_{J_1}$, the number of strings which are in the state with quantum
number $J_1$.  $J_1$ can have both integers and half integers values.   
The contribution of this state to total $J$ is $J_1
n_{J_1}$.  We enforce Bose statistics by summing over all occupation
numbers of all states, to get the partition function
\begin{eqnarray}
Z =e^{-\beta F}=\prod_{J=1/2,1,3/2\dots}^\infty \sum_{n_J=0}^\infty \exp \left(-
\frac{\beta  n_{J}J}{\alpha' f \sqrt{\lambda}}\right) =
\prod_{J=1/2,1,3/2\dots}^\infty \frac {1} 
{ 1- e^{ -\frac {\beta J} {\alpha' f \sqrt{\lambda}} } } 
\\
F=\frac{1}{\beta}\sum_{J=1/2,1,3/2\dots}^\infty\ln\left(1- 
e^{ -\frac {\beta J} {\alpha' f \sqrt{\lambda}} } \right)
=-\frac{1}{\beta}\sum_{p=1}^\infty\sum_{n=1}^\infty\frac{1}{n}
e^{ -\frac {n\beta p} {2\alpha' f \sqrt{\lambda}} } 
\\
=-\frac{1}{\beta}\sum_{n=1}^\infty\frac{1}{n}
\frac{1}{ e^{\frac {n\beta} {2\alpha' f \sqrt{\lambda}  } }-1 }
=-\frac{1}{\beta} \sum_{n=1}^\infty \frac{2\alpha' f\sqrt{\lambda}}{\beta n^2}
=-\frac{\pi^2\alpha' f\sqrt{\lambda}}{3\beta^2}
\label{ympartf}
\end{eqnarray}
where, $p=2J$ and, in the last step, we have taken the large $f$ limit. 
We will see that this coincides with the large $f$ limit of the string 
partition function which we will find in the following section. To do this, we 
need to identify the infinite length of the $X^-$-direction.  We can do this
by examining the quantization of $P^+$.  $J$ and $\Delta$ can be 
integers and, for fermions, 
half-integers, but in all cases, the sum $\Delta+J$ are integers.
Consequently, $P^+$ should be of the form
$\sqrt{2}\pi \cdot{\rm integer}/L$.  From this we identify the infinite length in the $X^9$-direction 
as $L=2\pi \alpha'f\sqrt{\lambda}$.   Then the large $f$ limit in (\ref{ympartf})
is
$$
F\to - \frac{\pi L}{6\beta^2}
$$

One could speculate about what happens when string interactions are switched on. 
In the asymptotically AdS space, which is dual to Yang-Mills theory with finite $J$, 
string interactions are restored by relaxing the large $N$ limit.  This produce a
cutoff of order $N^2$ on the number of independent traces of local operators, and therefore
it should cut off the Hagedorn behavior -- at least the counting of independent operators 
in weakly coupled
Yang-Mills theory no longer produces a Hagedorn spectrum.  Commensurate with this, we do not expect a 
de-confinement phase transition in Yang-Mills theory in 
the finite volume of $S^3\times S^1$ if $N$ is finite.  This makes the 
prediction that interacting strings do not have a finite temperature
phase transition on an asymptotically AdS space.

On the other hand, to obtain the plane-wave background, we should always take the limit $N\to\infty$.
This would suggest that we always have a Hagedorn spectrum of traces of local operators, the main
question being whether their quantum number $\Delta-J$ remains 
finite when both coupling constants, $\lambda'$ and $g_2$ are non-zero.  
It is known that when $\Delta-J$ depends on $g_2$ it is shifted by
a small amount when $g_2$ is small~\cite{Beisert:2002bb,Constable:2002vq}.
Thus, we can speculate that, as long as $g_2$ is small enough, the Hagedorn behavior indeed
persists when string interactions are present.   

\subsection{Other issues}

In the Yang-Mills partition function which is (\ref{partf}) with the
momenta (\ref{momenta}),  we should take $R/\sqrt{\alpha'}\to\infty$
while holding the temperature $\beta/\sqrt{\alpha'}$ fixed.  The
states which contribute in the trace are those which have finite
$\Delta-J$.  On the other hand, $\Delta+J$, and therefore  both
$\Delta$ and $J$    get arbitrarily large as
$R/\sqrt{\alpha'}\to\infty$.  One might question whether this limit is
sensible.  In the usual limit, $p^+$ and $p^-$ are held constant when
$N$ is taken to infinity.  Here, instead, the temperature is held
constant and it is not a priori clear that holding the temperature
constant and finite actually samples the states of the Yang-Mills
theory which coincide with the  string states.  It would be
interesting to find a way to check this directly.  Unfortunately, the
standard perturbative computation using path integrals is only valid
in the de-confined phase which occurs at high temperatures where the
confined states that we find in string theory would be difficult to
detect.
 
An important issue is the possible existence of zero modes of $p^+$.
Any protected operator for which $J^2/N\to 0$ as $N\to\infty$ are zero
modes of $p^+$.  Some of these are just at the $p^+=0$ edge of the
continuum spectrum and are included in our analysis.  These are the
operators  ${\rm Tr} Z^J$ where $J$ is not taken to infinity fast
enough as $N$ is taken to infinity.  There are also other operators,
such as the protected operators in the dilaton supermultiplet which
have finite non-zero $\Delta-J$ and for which  $\Delta+J$ are finite
in the limit as $N\to\infty$,  so that $p^+=0$. These could be
considered as discrete zero modes of $p^+$ which seem to have no
analog in the light-cone string spectrum.  This would seem to be a
miss-match between the string and Yang-Mills spectra.
 
Recently, there has been some discussion on the Hagedorn behavior
of pp-wave strings \cite{PandoZayas:2002hh,Greene:2002cd} also in 
discrete light cone quantization~\cite{Sugawara:2002rs}. 
It is well known that
when string theories are placed in a background electric NS $B$ field
or in a metric, the Hagedorn temperature depends on the parameters of the
background \cite{Grignani:2001hb,Grignani:2001ik,DeRisi:2002gt}.
Here we shall find that also the RR flux (\ref{5form})
felt by a string in the pp-wave metric 
modifies the Hagedorn temperature.
We shall also
clarify some of the issues related to the small and
large $f$ limit of the Hagedorn temperature, providing results
that even if in qualitative agreement with those of refs.
\cite{PandoZayas:2002hh,Greene:2002cd} differ quantitatively. 
We shall then study the
thermodynamic behavior of strings in geometries that arise in D1-D5
systems as $AdS_3\times S^3\times T^4$ with 
NS-NS and RR 3-form backgrounds~\cite{Russo:2002rq,Berenstein:2002jq,Sugawara:2002rs}.
It would be intersting to rederive our results 
by means of a path integral procedure and 
then generalize them to higher genera~\cite{Grignani:2000zm}.

\section{Free energy}

The free energy of a gas of noninteracting superstrings is given by summing the  
free energies of free particles over all of the particle species in the string spectrum
\footnote{ For a derivation of this formula, see 
ref.\cite{Grignani:1999sp}.}.  Each boson in the spectrum contributes
\be
F=\frac{1}{\beta}{\rm Tr} \ln\left( 1-e^{-\beta p^0}\right)=
-\sum_{n=1}^{\infty}\frac{1}{ n\beta}\mbox{Tr} 
e^{-\frac{n\beta}{\sqrt{2}}(p^{+}-p^{-})}.
\label{free1b}
\ee
where $p^0$ and $p^\pm$ are the energy and light-cone momenta of the particle.
Similarly, each fermion contributes
\be
F=-\frac{1}{\beta}
{\rm Tr} \ln\left( 1+e^{-\beta p^0}\right)=
\sum_{n=1}^{\infty}\frac{(-1)^{n}}{ n\beta}\mbox{Tr} e^{-\frac{n\beta}{\sqrt{2}}(p^{+}-p^{-})}.
\label{free1f}
\ee
We emphasize that the trace in each case is over the spectrum of single particle states, 
rather than multi-particle states.
The total free energy is given by summing (\ref{free1b}) and (\ref{free1f})
over the particles which appear in the string spectrum.   
Because of supersymmetry, most of the string spectrum has paired fermionic and bosonic states, 
so for that we can take the average of the two expressions,
\be
F_{\rm susy}= 
-\sum_{n=1}^{\infty}\frac{1-(-1)^n}{2 n\beta}\mbox{Tr} 
e^{-\frac{n\beta}{\sqrt{2}}(p^{+}-p^{-})}.
\label{free1susy}
\ee
However, there is one set of states in the spectrum which will turn out not
to have a superpartner, they are the lowest energy excitations which are
bosons and have vanishing light-cone Hamiltonian $p^-$ and arbitrary $p^+$.
To take these bosons into account (\ref{free1susy}) must be amended to read
\be
F= 
-\sum_{n=1,{\rm odd}}^{\infty}\frac{1}{ n\beta}\mbox{Tr}_{(p^-< 0)} 
e^{-\frac{n\beta}{\sqrt{2}}(p^{+}-p^{-})} 
-\sum_{n=1}^{\infty}\frac{1}{ n\beta}\mbox{Tr}_{(p^-= 0)} 
e^{-\frac{n\beta}{\sqrt{2}}(p^{+}-p^{-})}
\label{freetotal}
\ee
Summing these operators over the spectrum of the operators $p^-$ and $p^+$ which are  
found in light-cone quantization of the string should yield the free energy. The last
term is easily evaluated by noting that the measure for the trace over $p^+$ is 
$\frac{L}{\sqrt{2}\pi}\int_0^{\infty}
dp^+$, where $L$ is the (infinite) length of the $9$'th dimension.  We combine
the odd integer sum in the last term with the first term. This removes the
constraint on the spectrum in that term.  Then, 
\be
F=-\sum_{n=1,{\rm odd}}^{\infty}\frac{1}{ n\beta}\mbox{Tr}
e^{-\frac{n\beta}{\sqrt{2}}(p^{+}-p^{-})} -\frac{ L}{\pi\beta^2} 
\sum_{n=2,{\rm even}}^\infty
\frac{1}{n^2} =-\sum_{n=1,{\rm odd}}^{\infty}\frac{1}{ n\beta}\mbox{Tr}
e^{-\frac{n\beta}{\sqrt{2}}(p^{+}-p^{-})} -\frac{ L\pi}{24\beta^2} 
\label{free1total}
\ee 
To proceed, we must examine the string spectrum.

The Green-Schwarz IIB-superstring can be quantized in the light-cone gauge. 
The explicit form of the light-cone  Hamiltonian is 
\bea 
H&\equiv& -P^{-}\cr
&=&f(a_{0}^{I}\bar{a}_{0}^{I}+2\bar{\theta}_{0}\bar{\gamma}^{-}\Pi\theta_{0}+4)
+\frac{1}{\alpha'
p^{+}}\sum_{\mathcal{I}=1,2}\sum_{m=1}^{\infty}\sqrt{m^{2}+(\alpha'
p^{+}f)^{2}}(a_m^{\cI I} \bar{a}_m^{\cI I}+
\eta_m^\cI\bar{\gamma}^-\bar{\eta}_m^\cI )\cr
&=&f(N_{0}^{B}+N^{F}_{0}+4)+\frac{1}{\alpha'
p^{+}}\sum_{\mathcal{I}=1}^2\sum_{m=1}^{\infty}\sqrt{m^{2}+(\alpha'
p^{+}f)^{2}}(N_{\mathcal{I}m}^{B}+N_{\mathcal{I}m}^{F}).
\eea 
where we refer to \cite{Metsaev:2002re} for the notation.
 
The level matching condition $N_{1}=N_{2}$ also has to be enforced
by introducing an integration over the Lagrange multiplier $\tau_{1}$. 
Explicitly (\ref{free1total}) means
\bea
F&=&-\sum_{n=1,{\rm odd}}^{\infty}\frac{L}{4\pi^2\alpha'} \int_{0}^{\infty} 
\frac{d\tau_2}{\tau_2^2}\prod_{I=1}^8
\sum_{N_{0}^{B,I}=0}^{\infty}\sum_{n_{R},n_{L}=0}^{4}
\sum_{N_{1,2m}^{B,I}=0}^{\infty}\sum_{N_{1,2m}^{F}=0}^{8}\cr
&&\int_{-\frac{1}{2}}^{\frac{1}{2}}d\tau_{1}e^{2\pi i
\tau_{1}\sum_{m=1}^{\infty}m(N_{1m}^{B,I}+N_{1m}^{F}
-N_{2m}^{B,I}-N_{2m}^{F})}\cr
&& e^{-\frac{n^2\beta^2}{4\pi\alpha'\tau_2}}e^{-\frac{n\beta f
N_{0}^{B,I}}{\sqrt{2}}} \left(\begin{array}{c} 4\\
n_{R}\end{array}\right) \left(\begin{array}{c} 4\\
n_{L}\end{array}\right)e^{-\frac{n\beta
f}{\sqrt{2}}(-n_{R}+n_{L}+4)}\cr 
&& \left(\begin{array}{c} 8\\
N_{1m}^{F}\end{array}\right) \left(\begin{array}{c} 8\\
N_{2m}^{F}\end{array}\right)
e^{-\sum_{m=1}^{\infty}R_{m}(N_{1m}^{B,I}
+N_{1m}^{F}+N_{2m}^{B,I}+N_{2m}^{F})}
-\frac{ L\pi}{24\beta^2} 
\eea
where $L$ is the length of the longitudinal direction,
$N_{0}^{F}=-n_{R}+n_{L}$ and 
\be
R_{m}=2\pi\tau_{2}\sqrt{m^{2}+\mu^2}~,~~~
\tau_{2}=\frac{n\beta}{2\sqrt{2}\pi\alpha' p^{+}}~,~~~
\mu=\alpha'
p^{+}f=\frac{n\beta f}{2\sqrt{2}\pi\tau_{2}}
\label{mu}
\ee 
Due to the anticommutation relations of the creation-annihilation fermion operators,
the degeneracy of a state with $n_{\R,\L}$ fermions
is given by the binomial coefficient  $\left(\begin{array}{c} 4\\
n_{\R,\L}\end{array}\right)$.
Analogously the occupation number 
$N_{\mathcal{I}m}^{F}$ for the fermion non-zero modes,
which have eight independent components, runs from 0 to 8
and the degeneracy is given by the binomial coefficient  
$\left(\begin{array}{c} 8\\
N_{\mathcal{I}m}^{F}\end{array}\right)$.
Summing over the zero-modes, the free energy can be
written as 
\bea
F&=&-\sum_{n=1,{\rm odd}}^{\infty}\frac{L}{4\pi^2\alpha'} \int_{0}^{\infty}
\frac{d\tau_2}{\tau_2^2}
\int_{-\frac{1}{2}}^{\frac{1}{2}}d\tau_{1}e^{-\frac{n^2\beta^2
}{4\pi\alpha'\tau_2}} \left(\frac{1}{1-e^{-\frac{n\beta
f}{\sqrt{2}}}}\right)^{8}\cr 
&&e^{-\frac{4n\beta
f}{\sqrt{2}}}\left(1+e^{-\frac{n\beta
f}{\sqrt{2}}}\right)^{4}\left(1+e^{\frac{n\beta
f}{\sqrt{2}}}\right)^{4}|G(\tau_1,\tau_2,\frac{n\beta f}{\sqrt{2}2\pi\tau_2})|^{2}
-\frac{ L\pi}{24\beta^2} 
\label{free}
\eea 
$G$ is the function
\be
G(\tau_1,\tau_2,\mu)= \prod^8_{I=1}
\sum_{N_{1m}^{B,I}=0}^{\infty}\sum_{N_{1m}^{F}=0}^{8}\left(\begin{array}{c}
8\\ N_{1m}^{F}\end{array}\right)
e^{2\pi i
\tau_{1}\sum_{m=1}^{\infty}m(N_{1m}^{B,I}+N_{1m}^{F})}
e^{-\sum_{m=1}^{\infty}R_{m}(N_{1m}^{B,I}+N_{1m}^{F})}
\ee
Performing the sums over the occupation numbers 
the generating function becomes 
\be
G(\tau_1,\tau_2,\mu)=
\prod_{m=1}^{\infty}\left(
\frac{1+e^{-2\pi\tau_{2}\sqrt{m^{2}+\mu^{2}}
+2\pi i\tau_1 m}}
{1-e^{-2\pi\tau_{2}\sqrt{m^{2}+\mu^{2}}
+2\pi i\tau_1 m}}\right)^{8}
\label{G}
\ee
so that the free energy reads
\be
F=-\sum_{n=1,{\rm odd}}^{\infty}\frac{L}{4\pi^2\alpha'} \int_{0}^{\infty}
\frac{d\tau_2}{\tau_2^2}
\int_{-\frac{1}{2}}^{\frac{1}{2}}d\tau_{1}e^{-\frac{n^2\beta^2
}{4\pi\alpha'\tau_2}}
\prod_{m=-\infty}^{\infty}\left(
\frac{1+e^{-2\pi\tau_{2}\sqrt{m^{2}+\mu^{2}}
+2\pi i\tau_1 m}}
{1-e^{-2\pi\tau_{2}\sqrt{m^{2}+\mu^{2}}
+2\pi i\tau_1 m}}\right)^{8}
-\frac{ L\pi}{24\beta^2} 
\label{freev}
\ee 
This equation is {\bf not} in agreement with eq.(3.3) of ref. \cite{PandoZayas:2002hh},
because it differs by the contribution of the zero light-cone energy mode.
The limit $f\to\infty$ of (\ref{freev}) can be easily computed 
since $G(\tau_1,\tau_2,\mu)\to 1$ in this limit so that $F$ becomes
\be
F=-\frac{\pi L}{6 \beta^2}
\ee
which 
coincides with the free energy of the dual gauge theory (\ref{ympartf})
computed in this limit in the introduction.  Note that this is the
free energy density of a gas of massless particles in two dimensions.
Indeed, the lowest energy states of the string are massless chiral bosons
which propagate down the two spacetime dimensional axis of the pp-wave
space made of the $X^+$ and $X^-$ directions.  
Further they are chiral, in that their spectrum is composed 
entirely of left-moving particles.  The spectrum of these particles is protected by supersymmetry, 
so we expect that this limit of the partition function is
not corrected by string interactions.

We shall now extract information directly from (\ref{freev}) 
instead of turning
to the path integral approach as in \cite{PandoZayas:2002hh}.
To compute the Hagedorn temperature we need to estimate the asymptotic 
behavior of the product in (\ref{freev}). This will be crudely estimated in 
this section;
a more precise estimate will be obtained in the next section 
by using its modular transformations properties \cite{Takayanagi:2002pi}.
Consider the function defined by
\be
Z(\tau_1,\tau_{2},\mu)\equiv\prod_{m=-\infty}^{\infty}\left(
\frac{1+e^{-2\pi\tau_{2}\sqrt{m^{2}+\mu^{2}}
+2\pi i\tau_1 m}}
{1-e^{-2\pi\tau_{2}\sqrt{m^{2}+\mu^{2}}
+2\pi i\tau_1 m}}\right)
\label{etapp}
\ee 
It diverges only when
$\tau_1$, $\tau_2$ and $\beta f$ vanish, let us then consider
these limits by taking first $\tau_1=0$ and then $\tau_2\to 0$.
For $\tau_1=0$ it reads
\bea
Z(0,\tau_{2},\mu)&=&
\exp\left\{\sum_{m=-\infty}^{\infty}
\ln\left(
\frac{1+e^{-2\pi\tau_{2}\sqrt{m^{2}+\mu^{2}}}}
{1-e^{-2\pi\tau_{2}\sqrt{m^{2}+\mu^{2}}}}\right)\right\}\cr
&=&\exp\left\{-\sum_{m=-\infty}^{\infty}\sum^{\infty}_{p=1}
\left[\frac{(-1)^{p}}{p}-\frac{1}{p}\right]
e^{-2\pi \tau_2p\sqrt{m^{2}
+\mu^2}}\right\}
\eea

Using the integral identity \cite{gradsteyn}
\be
e^{-2\sqrt{ab}}\frac{1}{2}\sqrt{\frac{\pi}{a}}
=\int_{0}^{\infty}e^{-at^{2}-\frac{b}{t^{2}}}dt,\qquad
(a,b>0)
\ee 
one can write 
\be
Z=\exp\left\{+2\sum_{m=-\infty}^{\infty}
\sum^{\infty}_{p_{odd}=1}\frac{1}{p}
\frac{2}{\sqrt{\pi}}\int_{0}^{\infty}
e^{-t^{2}-\frac{\pi^{2}\tau^{2}_{2}p^{2}m^{2}}{t^{2}}
-\frac{n^{2}\beta^{2}f^{2}p^{2}}{8t^{2}}}dt\right\}
\ee
In the limit of interest $\tau_{2}\rightarrow 0$, the sum over $m$ may
be approximated by an integral $\sum_{m}\simeq\int_{-\infty}^{\infty}dm$.
The integration over $m$ is gaussian and can be performed. 
The leading behavior in the $\tau_2\to 0$ limit then is
\be
Z\simeq\exp\left\{\frac{n\beta
f}{\sqrt{2}\pi\tau_{2}}\sum^{\infty}_{p=1}\frac{[1-(-1)^p]}{p}K_{1}
\left(\frac{n\beta f
p}{\sqrt{2}}\right)\right\}
\label{Z}
\ee 
where $K_{1}$ is the modified Bessel function. Using the series expansion of
the Bessel function it is easy to see that the leading term in the limit
 $\beta f\to 0$ (\ref{Z}) reproduces the expected flat space behavior.
A more precise derivation of this result will be obtained in the
next section.

\section{Modular properties of $Z$}

Consider the function defined by 
\be
Z_{a,b}(\tau_1,\tau_2,x)=
\prod_{m=-\infty}^{\infty}(1-e^{-2\pi\tau_2\sqrt{x^2+(m+b)^2}
+2\pi i\tau_1(m+b)+2\pi ia}).
\label{zetaab}
\ee
The partition function (\ref{etapp}) is given by the ratio
\be
Z(\tau_1,\tau_2,\frac{n\beta f}{\sqrt{2}2\pi\tau_2})=
\frac{Z_{\frac{1}{2},0}(\tau_1,\tau_2,\frac{n\beta f}{2\pi\sqrt{2}\tau_2})}
{Z_{0,0}(\tau_1,\tau_2,\frac{n\beta f}{2\pi\sqrt{2}\tau_2})}
\label{z}
\ee
It will turn out to be  useful to define
\be
\Delta_b(x)=-\frac{1}{2\pi^2}\sum_{p=1}^{\infty}\cos(2\pi bp)
\int^{\infty}_{0} ds\ 
e^{-p^2s-\frac{\pi^2 x^2}{s}}=
-\frac{x}{\pi}\sum_{p=1}^{\infty}\frac{\cos(2\pi b p)}{p}
K_{1}\left(2 \pi x p\right)
\label{casimir}
\ee
The quantity $\Delta_b(x)$ corresponds to 
the zero-energy (Casimir energy)
of a 2D complex scalar boson $\phi$ of mass $m$ with the 
twisted boundary condition 
$\phi(\tau,\sigma+\pi)=e^{2\pi ib}\phi(\tau,\sigma)$.
In the massless limit this zero energy correctly reproduces the familiar
value
\be
\lim_{x\to 0}\Delta_{b}(x)=\frac{1}{24}-\frac{1}{8}(2b-1)^2.
\ee
Following the appendix A of ref. \cite{Takayanagi:2002pi} it is not difficult
to derive the modular property of (\ref{zetaab})
\be
\ln Z_{a,b}(\tau_1,\tau_2,x)=\ln Z_{-b,a}\left(-\frac{\tau_1}{|\tau|^2},
\frac{\tau_2}{|\tau|^2},\frac{x}{|\tau|}\right)
-2\pi\tau_2 \Delta_{b}(x)+2\pi
\frac{\tau_2}{|\tau|^2} 
\Delta_{a}\left(\frac{x}{|\tau|}\right)
\label{modab}
\ee
As a consequence the transformation properties of (\ref{etapp}) are
\bea
\ln Z\left(\tau_1,\tau_2,\frac{n\beta f}{\sqrt{2}2\pi\tau_2}\right)&=&
\ln Z_{0,\frac{1}{2}}\left(-\frac{\tau_1}{|\tau|^2},
\frac{\tau_2}{|\tau|^2},\frac{n\beta f|\tau|}{2\pi\sqrt{2}\tau_2}\right)
-\ln Z_{0,0}\left(-\frac{\tau_1}{|\tau|^2},
\frac{\tau_2}{|\tau|^2},\frac{n\beta f|\tau|}{2\pi\sqrt{2}\tau_2}\right)\cr
&+&2\pi\frac{\tau_2}{|\tau|^2}\left[ 
\Delta_{\frac{1}{2}}\left(\frac{n\beta f|\tau|}{2\pi\sqrt{2}\tau_2}\right)
-\Delta_{0}\left(\frac{n\beta f|\tau|}{2\pi\sqrt{2}\tau_2}\right)\right]
\label{modular}
\eea
From the definition of the Casimir energy (\ref{casimir}) the last two 
terms in (\ref{modular}) read
\be
\frac{n\beta f}{\sqrt{2}\pi|\tau|}\sum^{\infty}_{p=1}
\frac{[1-(-1)^p]}{p}K_{1}\left(\frac{n\beta f
p|\tau|}{\sqrt{2}\tau_2}\right)
\ee
In the limit $\tau_1\to 0$ and $\tau_2\to 0$ the first two terms
in (\ref{modular}) behave smoothly whereas the second two give precisely the 
behavior found in (\ref{Z}).

\section{Hagedorn temperature}

The asymptotic value of the free energy 
(\ref{free}) then is
\be
F\sim\sum_{n=1}^{\infty}\frac{(-1)^{n}-1}
{8\pi^2\alpha'}L \int_{0}^{\infty}\frac{d\tau_2}{\tau_2^2}
e^{-\frac{n^2\beta^2}{4\pi\alpha'\tau_2}}
\exp\left\{\frac{8 n\beta
f}{\sqrt{2}\pi\tau_2}\sum^{\infty}_{p=1}\frac{[1-(-1)^p]}{p}K_{1}
\left(\frac{n\beta f
p}{\sqrt{2}}\right)\right\}
\label{asymf}
\ee
The biggest value of $\beta$ for which this expression diverges in the 
$\tau_2\to 0$ limit is obtained by 
taking the $n=1$ mode. 
When the exponent in the integrand of (\ref{asymf}) vanishes, $F$ 
starts to diverge  
so that the Hagedorn temperature is defined by the equation
\be
\frac{\beta_H^2}{4\pi\alpha'}=\frac{8 \beta_H
f}{\sqrt{2}\pi}\sum^{\infty}_{p=1}
\frac{[1-(-1)^p]}{p}K_{1}\left(\frac{\beta_H f p}{\sqrt{2}}\right)
\label{pphag}
\ee
Taking the derivative with respect to $f$ one gets
\be
\frac{\partial \beta_H}{\partial f}=
-\frac{8\alpha' |f|\beta_H \sum_{p=1}^{\infty}\left[1-(-1)^p\right]K_{0}\left(\frac{\beta_H f p}{\sqrt{2}}\right)}
{1+8\alpha' f^2\sum_{p=1}^{\infty}\left[1-(-1)^p\right]K_{0}\left(\frac{\beta_H f p}{\sqrt{2}}\right)}
\label{der}
\ee
The r.h.s. of this equation is always negative
thus $\beta_H$ is a decreasing function of $|f|\sqrt{\alpha'}$
and consequently $T_H$ is an increasing function
of $|f|\sqrt{\alpha'}$.

We shall now study the behavior of equation (\ref{pphag}) in
the small and large $f$ limit.
For small $f$
it is necessary to rewrite it as a power series in $\beta f$
and then solve for $\beta$. This will be rigorously done in the
the next section and it will allow us to 
derive the correct result for the Hagedorn 
temperature at small $f$. For large $f$ the behavior of
(\ref{pphag}) it is much easier to extract and it should
reproduce the dual gauge theory behavior.

\subsection{Expansion for small $f$}

To rewrite (\ref{pphag}) as a series expansion in $f$, 
we shall use the Mellin transform procedure.
The series 
\be
S_b(x)=\sum^{\infty}_{p=1}
\frac{1}{p}K_{1}\left(x p\right)
\ee
can in fact be rewritten as a power series in $x$ by means of a Mellin
transformation. The Mellin transform of $S_b(x)$ reads
\be
M(s)=\int_0^\infty dx x^{s-1} S_b(x)=\sum^{\infty}_{p=1}
\int_0^\infty dx \int_0^\infty \frac{dt}{4 t^2}  x^s
\ e^{-t-\frac{x^2 p^2}{4 t}}
\ee
Changing the integration variable $x$ to $y=x^2 p^2/(4 t)$,
$M(s)$ becomes
\be
M(s)=\sum^{\infty}_{p=1}
\int_0^\infty \frac{dy}{8}\left(\frac{2}{p}\right)^{s+1} y^{(s-1)/2} e^{-y}
\int_0^\infty \frac{dt}{t^2} t^{(s+1)/2} e^{-t}
\ee
The Mellin transform $M(s)$ exists 
provided the integrals over $y$ and $t$ are bounded for some $s>k$
with $k>0$. In our case the integrals can be done for $s>1$ and
$M(s)$ is
\be
M(s)=2^{s-2}\Gamma\left(\frac{s-1}{2}\right)
\Gamma\left(\frac{s+1}{2}\right) \zeta (s+1)
\ee
The inversion of the Mellin transform gives back the function $S_b(x)$
and is accomplished by means of the inversion integral
\be
S_b(x)=\frac{1}{2\pi i}\int^{C+i\infty}_{C-i\infty} ds
M(s) x^{-s}
\ee
where $C>k=1$.
The integral is well defined and to compute it we must close the contour
and use the residue theorem. For this purpose it is convenient
to change the argument of $\zeta(s+1)$ in the integrand as \cite{gradsteyn}
\be
\zeta(s+1)=\pi^{s+1/2}\frac{\Gamma\left(-\frac{s}{2}\right)}
{\Gamma\left(\frac{s+1}{2}\right)} \zeta (-s)
\ee
Therefore
\be
S_b(x)=\frac{1}{2\pi i}\int^{C+i\infty}_{C-i\infty} ds 
\left(\frac {2\pi}{x}\right)^{s}\frac{\sqrt{\pi}}{4}
\Gamma\left(-\frac{s}{2}\right)
\Gamma\left(\frac{s-1}{2}\right)
\zeta (-s)
\ee
The contour can now be closed on the left so that 
the poles are at $s=1,0,-1,1-2k,\dots$ for $k=2,3,\dots$. 
The residues can be easily computed and the result is
\bea
S_b(x)&=&\frac{\pi^2}{6 x}-\frac{\pi}{2}+ \frac{x}{8}
\left(1-2\gamma+2 \ln\frac{4\pi}{x}\right)\cr
&+&
\sum^\infty_{k=2}\frac{(-1)^k}{k!}
\left(\frac{x} {2\pi}\right)^{2k-1}\frac{\sqrt{\pi}}{2}
\Gamma\left(k-\frac{1}{2}\right)
\zeta (2k-1)
\label{besselb}
\eea
where $\gamma$ is the Euler constant.
Analogously one can rewrite the series
\be
S_f(x)=\sum^{\infty}_{p=1}
\frac{(-1)^p}{p}K_{1}\left(x p\right)
\ee
as
\bea
S_f(x)&=&-\frac{\pi^2}{12 x}+ \frac{x}{8}
\left(1-2\gamma+2 \ln\frac{\pi}{x}\right)\cr
&+&
\sum^\infty_{k=2}\frac{(-1)^k}{k!}(2^{2k-1}-1)
\left(\frac{x} {2\pi}\right)^{2k-1}\frac{\sqrt{\pi}}{2}
\Gamma\left(k-\frac{1}{2}\right)
\zeta (2k-1)
\eea
The series appearing in the formula for the Hagedorn temperature
(\ref{pphag}) can then be rewritten as
$$
S_b(x)-S_f(x)=\sum^{\infty}_{p=1}
\frac{1-(-1)^p}{p}K_{1}\left(x p\right)
$$
\be
=\frac{\pi^2}{4 x}-\frac{\pi}{2}+ \frac{x}{2}
\ln 2-
\sum^\infty_{k=2}\frac{(-1)^k}{k!}(2^{2k-1}-2)
\left(\frac{x} {2\pi}\right)^{2k-1}\frac{\sqrt{\pi}}{2}
\Gamma\left(k-\frac{1}{2}\right)
\zeta (2k-1)
\label{s}
\ee

Using these results for the series difference in (\ref{pphag})
one can derive the following formula for the Hagedorn temperature in the 
limit of small $f$.
\be
\frac{\beta_H^2}{4\pi\alpha'}
=2\pi-\frac{4 \beta_Hf}{\sqrt{2}}
+\frac{2\beta_H^{2}f^{2}\ln2}{\pi}
-\sum_{k=2}^{\infty}\frac{(-1)^k(2^{2k}-4)4\sqrt{\pi}}{k!}
\left(\frac{\beta_H f}{2\pi\sqrt{2}}\right)^{2k}
\Gamma(k-\frac{1}{2})\zeta(2k-1)
\label{pphagsf}
\nonumber
\ee
Keeping only the two leading terms in the expansion of (\ref{pphagsf}) we get
\be 
\beta_H^{2} ({1-8\alpha' f^{2}\ln2})=8\pi^2\alpha' - 
\frac{16\pi\alpha'\beta_H f}{\sqrt{2}}
\label{hagedorn}
\ee 
The Hagedorn temperature then is
\be 
T_H=\frac{1}{2\pi\sqrt{2\alpha'}}\left(1+2\sqrt{\alpha'}f
+2(1-2\ln 2)\alpha'f^2 \right)
\ee 
As in \cite{PandoZayas:2002hh} the Hagedorn temperature
increases for small values of $f^2\alpha'$ but the second
term differs from the one derived in \cite{PandoZayas:2002hh}
by a factor of $4\pi\sqrt{2}$.

In the flat space limit $f\rightarrow 0$ we recover the well known 
superstring Hagedorn temperature 
\be
T_{H}=\frac{1}{\beta_H}=\frac{1}{2\pi \sqrt{2\alpha'}}
\ee

\subsection{The large $f$ limit}

Let us now consider the large $f$ behavior of eq. (\ref{pphag}).
It is particularly interesting to examine this limit because it is
in this limit pp-wave that type-IIB string theory is supposed to
be dual to a subsector of a particular Yang-Mills 
theory~\cite{Berenstein:2002jq}.
For large value of $f$, the most  relevant contribution to the series 
of the modified Bessel function $K_{1}$ is given by taking $p=1$
in (\ref{pphag}).
For large value of its argument the Bessel function in fact can be
approximated by
\be
K_{1}\left(\frac{\beta_H f p}{\sqrt{2}}\right)
\sim \sqrt{\frac{\pi}{\sqrt{2}\beta_H f p}}\exp 
\left(-\frac{\beta_H f p}{\sqrt{2}}\right)
\label{asympt}
\ee
so that terms with higher values of $p$ are exponentially suppressed.
The equation (\ref{pphag}) for the Hagedorn temperature becomes
\be
\frac{\beta_H^2}{4\pi\alpha'}=
8\sqrt{\frac{\beta_H f \sqrt{2}}{\pi}}
\exp \left(-\frac{\beta_H f}{\sqrt{2}}\right)
\to_{f\to \infty }0
\label{aspphag}
\ee
The rapid vanishing of the Bessel function in the large $f$ limit
implies that the Hagedorn temperature increases with $f$ and
for very large $f$ is pushed toward infinity. 
This means that in this regime there is no Hagedorn transition
at any finite temperature but instead the Hagedorn temperature
is a limiting temperature. This is expected since
the large $f$ limit should indeed reproduce the gauge theory 
behavior.

\section{$AdS_3 \times S^3$ in NS-NS and RR $3$-form backgrounds}

The limit that gives the metric (\ref{metric}) in the $AdS_5 \times S^5$
geometry can be taken also in other geometries. As a particular case
one can consider the $AdS_3 \times S^3$ geometry
\cite{Nappi:1993ie,Russo:2002rq,Berenstein:2002jq}.  In this case
the radii of $AdS_3$ and $S^3$ are the same and the computation
is identical to the one we did above for  $AdS_5 \times S^5$.
It is interesting to consider a situation with a mixture of NS-NS
and RR $3$-form field strengths.  The six dimensional plane-wave metric is
\be
ds^2=2dx^{+}dx^{-}-f^{2}{\vec y}^{2}dx^{+}dx^{+}+d{\vec y}^{2}
\label{ads3}
\ee
$$ 
F_{+12}^{NS}=F_{+34}^{NS}=C_1f\cos \alpha 
$$
\be
F_{+12}^{RR}=F_{+34}^{R}=C_2f\sin \alpha
\label{3form}
\ee
where ${\vec y}$ parametrizes a point on $T^4$ and $\alpha $
is a fixed parameter which allows us to interpolate between
the purely NS background $\alpha =0$ and the purely RR 
background $\alpha =\pi /2$. $C_1$ and $C_2$ are constants depending
on the string coupling and the normalization of the NS and RR 
field strenghts. In addition to the six coordinates in (\ref{ads3})
we have four additional directions which we can take to be $T^4$.

The light-cone Hamiltonian is
\be
H=\sum_{n=-\infty}^{\infty}N_n \sqrt{f^2 {\sin{\alpha}}^2
+\left(f\cos \alpha +\frac{n}{\alpha ' p^+}\right)^2} +
2\frac{L_0^{T^4} +{\bar L_0^{T^4}}}{\alpha ' p^+}
\label{hads3}
\ee
where the first term takes into account the massive bosons
and fermions and the second term takes into account the
massless bosons and fermions.

The computation of the free energy is similar to the one we performed
in the previous section for the $AdS_5 \times S^5$ geometry. 
It reads
\bea
&&F=-2^5\sum_{n_{odd}}^{\infty}\frac{L}{8\pi^2\alpha'} \int_{0}^{\infty}
\frac{d\tau_2}{\tau_2^2}
\int_{-\frac{1}{2}}^{\frac{1}{2}}d\tau_{1}
\left(\frac{1}{4\pi^2\alpha'\tau_2}\right)^2
e^{-\frac{n^2\beta^2
}{4\pi\alpha'\tau_2}}\prod_{m=1}^{\infty}\left|\frac{1+e^{2\pi i\tau m}}
{1-e^{2\pi i\tau m}}\right|^8\cr
&&\prod_{m=-\infty}^{\infty}\left\{
\frac{1+\exp\left[-2\pi\tau_2 
\sqrt{\left(\frac{n f\beta\sin\alpha}{2\sqrt{2}\pi\tau_2}\right)^2+
\left(m+\frac{n\beta f\cos\alpha}{2\sqrt{2}\pi\tau_2}\right)^2}
+2\pi i\tau_1 m\right]}
{1-\exp\left[-2\pi\tau_2 
\sqrt{\left(\frac{n f\beta\sin\alpha}{2\sqrt{2}\pi\tau_2}\right)^2+
\left(m+\frac{n\beta f\cos\alpha}{2\sqrt{2}\pi\tau_2}\right)^2}
+2\pi i\tau_1 m\right]}\right\}^{4}
\label{freeads3}
\eea
Here, because of the fermion zero modes, the ground state is degenerate
and the free energy can be computed using $F_{\rm susy}$ defined in 
equation (\ref{free1susy}).

The modular properties of the partition function in (\ref{freeads3}) 
can be derived as in section 3. Consider
\be
Z(\tau_1,\tau_2,x)=
\prod_{m=-\infty}^{\infty}\left(\frac{1+e^{-2\pi\tau_2\sqrt{x^2+(m+b)^2}
+2\pi i\tau_1(m+b)+2\pi ia}}{1-e^{-2\pi\tau_2\sqrt{x^2+(m+b)^2}
+2\pi i\tau_1(m+b)+2\pi ia}}\right)\left|\theta_4(0,2\tau)\right|^{-2}
\label{zetaads3}
\ee
In our case
$a=0$, $b=\frac{n \beta  f\cos \alpha}{2\sqrt{2}\pi\tau_2}$ 
and $x=\frac{n \beta  f\sin \alpha}{2\sqrt{2}\pi\tau_2}$.
(\ref{zetaads3}) can be rewritten in terms of the definition 
(\ref{zetaab}) as 
\be
Z(\tau_1,\tau_2,x)=
\frac{Z_{\frac{1}{2},b}(\tau_1,\tau_2,x)}
{Z_{0,b}(\tau_1,\tau_2,x)}\left|\theta_4(0,2\tau)\right|^{-2}
\ee
From the modular property of $Z_{a,b}(\tau_1,\tau_2,x)$,
eq. (\ref{modab}), it follows that
\bea
&\ln Z\left(\tau_1,\tau_2,\frac{n\beta f\sin \alpha }{\sqrt{2}2\pi\tau_2}\right)=
\ln Z_{-b,\frac{1}{2}}\left(-\frac{\tau_1}{|\tau|^2},
\frac{\tau_2}{|\tau|^2},\frac{n\beta f\sin \alpha |\tau|}{2\pi\sqrt{2}\tau_2}\right)
-\ln Z_{-b,0}\left(-\frac{\tau_1}{|\tau|^2},
\frac{\tau_2}{|\tau|^2},\frac{n\beta \sin \alpha f|\tau|}{2\pi\sqrt{2}\tau_2}\right)
\cr&+2\pi\frac{\tau_2}{|\tau|^2}\left[ 
\Delta_{\frac{1}{2}}\left(\frac{n\beta f\sin \alpha |\tau|}{2\pi\sqrt{2}\tau_2}\right)
-\Delta_{0}\left(\frac{n\beta f\sin \alpha |\tau|}{2\pi\sqrt{2}\tau_2}\right)\right]
-2\ln\left|\theta_2(0,-\frac{1}{2\tau})\right|+\ln2|\tau|
\label{modularads3}
\eea
The first two terms in (\ref{modularads3}) behave smoothly in the
$\tau_1\to 0$, $\tau_2\to 0$ limit. Moreover
$$\left|\theta_2(0,-\frac{1}{2\tau})\right|\to 
\exp\left(-\frac{\pi\tau_2}{4|\tau|^2}\right)$$
Consequently, taking into account the definition of the Casimir energies, 
for the Hagedorn temperature we get
\be
\frac{\beta_H^2}{4\pi\alpha'}=\frac{4\beta_H
 f\sin \alpha }{\sqrt{2}\pi}\sum^{\infty}_{p=1}
\frac{[1-(-1)^p]}{p}K_{1}\left(\frac{p \beta_H  f \sin \alpha}{\sqrt{2}}\right)
+\pi
\label{pphagads3}
\ee
It is interesting to note that this equation depends
on the angle $\alpha$ only through $f\sin \alpha$, the RR field
strenght \footnote{We thank Y. Sugawara
for pointing out the misprint present in the previous version of the Paper.}.

Keeping only the two leading terms in the expansion for small $ f$
of (\ref{pphagads3}) we get the Hagedorn temperature 
\be 
T_H=\frac{1}{2\pi\sqrt{2\alpha'}}\left(1+\sqrt{\alpha'} f\sin \alpha
+(1-2\ln 2)\alpha' f^2 \sin^2 \alpha \right)
\ee 
In the case of purely NS background, corresponding to $\alpha =0$,
we recover the well known  superstring Hagedorn temperature for the
flat background.

\acknowledgments 
The authors acknowledge Emil Akhmedov, Laura Fagiolini, Shiraz Minwalla, Giuseppe Nardelli,
Moshe Rozali,
Kristen Schleich, Bill Unruh and Mark Van Raamsdonk for helpful discussions.
G.W.S. acknowledges the Aspen Center for Physics where part of this
work was completed.

\appendix{\section{Alternative derivation of eq.(4.14)}}

The same expression for the difference $S_b(x)-S_f(x)$ can be obtained
using a completely different procedure. 
The series $S_b(x)-S_f(x)$ can in fact 
be obtained also from the formula
\be
S_b(x)-S_f(x)=-\frac{d}{dx}\left(x^2 \int_0^{\pi /x}dt' 
\int_0^{t'}dt \sum^{\infty}_{p=1}K_0\left(x p\right)\cos pxt \right)
\label{s1}
\ee
Using the fact that \cite{gradsteyn}
$$
\sum^{\infty}_{p=1}K_0\left(x p\right)\cos pxt = \frac{1}{2}
\left(\gamma + \ln \frac{x}{4 \pi} \right) +
\frac{\pi}{2x\sqrt{1+t^2}} 
$$
\be
+\frac{\pi}{2}\sum^{\infty}_{l=1}\left\{
\frac{1}{\sqrt{x^2 +(2l \pi - tx)^2}}-\frac{1}{2l \pi}\right\}
+\frac{\pi}{2}\sum^{\infty}_{l=1}\left\{
\frac{1}{\sqrt{x^2 +(2l \pi + tx)^2}}-\frac{1}{2l \pi}\right\}
\ee
equation (\ref{s1}) becomes
\bea
&&S_b(x)-S_f(x)=-\frac{\pi ^2}{4x} -\frac{\pi}{2}
+\frac{\pi}{2}\left(\frac{\sqrt{x^2 +\pi ^2}}{x} +
\frac{x}{\pi +\sqrt{x^2 +\pi ^2}}\right)
\cr
&&-\pi x \sum^{\infty}_{l=1}\left(\frac{1}{2l \pi +\sqrt{x^2 +(2l \pi)^2}}
-\frac{1}{(2l+1) \pi +\sqrt{x^2 +(2l+1)^2 \pi ^2}}\right)
\cr
&=&\frac{\pi ^2}{4x}+\frac{\pi}{2} -\pi x \sum^{\infty}_{l=0}
\left(\frac{1}{2l \pi +\sqrt{x^2 +(2l \pi)^2}}
-\frac{1}{(2l+1) \pi +\sqrt{x^2 +(2l+1)^2 \pi ^2}}\right)
\cr
&=&\frac{\pi ^2}{4x}-\frac{\pi}{2}-\pi x\sum^{\infty}_{k=1}
\frac{(-1)^k}{\pi k +\sqrt{x^2 +\pi ^2 k^2}}
\label{s2}
\eea
Expanding for small values of $x$, 
it is easy to prove that equation (\ref{s2}) 
becomes precisely equation (\ref{s})
\be
S_b(x)-S_f(x)=\frac{\pi ^2}{4x} -\frac{\pi}{2}
+\pi \sum^{\infty}_{k=1}\frac{(-1)^{k}}{k!\sqrt \pi}
\Gamma\left(k-\frac{1}{2}\right)
\left(\frac{x}{\pi }\right)^{2k-1}T_{2k-1} 
\label{s3}
\ee
where $T_{s}=\left(2^{1-s}-1\right)\zeta (s)$ and $T_1 =-\ln 2$
\cite{prudnikov}.

\end{document}